\documentstyle[12pt,aasms4]{article}
\begin{document}

\title{Origins of the slow and the ubiquitous fast solar wind}

\author{S. R. Habbal$^1$, R. Woo$^{2}$, S. Fineschi$^1$, 
R. O'Neal$^1$, J. Kohl$^1$,  G. Noci$^{3}$ and C. Korendyke$^{4}$}

\affil{\it $^1$Harvard-Smithsonian Center for Astrophysics,
Cambridge, MA 02138, USA}
\authoremail{shabbal$@$cfa.harvard.edu,
sfineschi$@$cfa.harvard.edu, roneal$@$cfa.harvard.edu,
jkohl$@$cfa.harvard.edu}
\affil{\it $^{2}$ Jet Propulsion Laboratory, California
Institute of Technology, Pasadena, CA 91109, USA}
\authoremail{richard@oberon.jpl.nasa.gov}
\affil{\it $^{3}$ Universita di Firenze, I-50125 Firenze, Italy}
\authoremail{noci@arcetri.astro.it}
\affil{\it $^{4}$ E. O. Hulburt Center for Space Research,
Naval Research Laboratory, Washington, D.C., 20375}
\authoremail{koren@cyclops.nrl.navy.mil}

\begin{abstract}

We present in this Letter the first coordinated radio occultation
measurements and ultraviolet observations of the inner corona below
5.5 $\rm R_s$, obtained during the Galileo solar conjunction in January 1997,
to establish the origin of the slow solar wind.
Limits on the flow speed are derived from the Doppler dimming of
the resonantly scattered component of the oxygen 1032 \AA\ and 1037 \AA\
lines as measured with the UltraViolet Coronagraph Spectrometer (UVCS) on
the Solar and Heliospheric Observatory (SOHO).
White light images of the corona from
the Large Angle Spectroscopic Coronagraph (LASCO) on SOHO taken
simultaneously are used to place the Doppler radio scintillation
and ultraviolet measurements in the context of coronal
structures. These combined observations provide
the first direct confirmation of the view recently proposed by Woo and
Martin (1997) that the slow solar wind is associated with the
axes, also known as stalks, of streamers. 
Furthermore, the ultraviolet observations
also show how the fast solar
wind is ubiquitous in the inner corona, and that a velocity shear
between the fast and slow solar wind develops along
the streamer stalks.
\end{abstract}

\keywords{Sun - Corona, Sun - Solar wind}

\section{Introduction}

The solar wind is a direct manifestation of the coronal heating
processes which continue to elude us. For over three decades,
observations in interplanetary space have identified
two types of wind: a slow component with highly variable physical
properties also characterized by speeds typically below
500 km/s, and a much less variable fast wind flowing on average at
750  km/s (e.g., Schwenn 1990; Gosling et al. 1995; Phillips et al. 1995).
Connecting these two types of winds to their origins at the
Sun is still not resolved. The prevailing view
is that the fast solar wind observed in the ecliptic plane originates from
the equatorial extensions of polar coronal holes onto the solar disk,
and occasionally from equatorial coronal holes
(e.g., Krieger, Timothy, $\&$ Roelof 1973; Bell and Noci 1976;
Hundhausen 1977). In addition, 
since the  boundaries of polar coronal hole do not extend below 50\arcdeg\
latitude even at solar minimum (e.g., Harvey 1996),
a faster than radial expansion of the magnetic field lines
originating from the polar caps (e.g., Munro and Jackson 1977)
has been recently invoked to account for the fast wind observed
out of the ecliptic by Ulysses from the poles down to about 30\arcdeg\
latitude (e.g., Gosling et al. 1995).
The source of the slow solar wind, on the other hand, generally believed to
be associated with the highly structured and variable streamer belt
has remained more enigmatic
(see, e.g., reviews by Schwenn 1990; Gosling 1997).

Important clues regarding the association of fast and slow solar wind with
the corresponding magnetic field structures in the corona have
recently emerged from remote sensing radio
occultation measurements. Large-scale gradients in velocity 
indicated that the slow wind emanated from localized sources in the
corona overlying the streamer belt (Woo 1995). 
That the slowest wind coincided with conspicuously high levels of
density fluctuation characteristic of coronal streamer stalks
(Woo et al. 1995) provided the first observational evidence that
the streamer stalks were the sources of the slow solar wind 
(Woo $\&$ Martin 1997).
(We refer to stalks as the narrowing of the streamers that extend into
interplanetary space, as evident in solar eclipse pictures (Koutchmy 1977).)
On the other hand, 
low levels of density fluctuations were found to
be characteristic of the fast solar wind (Woo $\&$ Gazis
1993; Woo $\&$ Martin 1997), suggesting that
they could be used as a proxy for the fast wind.
By comparing radio ranging measurements
with simultaneous white light observations, Woo and Habbal
(1997) found that low levels of density fluctuations could be traced back to
coronal holes as well as to quiet Sun regions (i.e. the diffuse
emission seen above $10^6$ K, filling the space between polar coronal
holes and the active region belt). Taken together with the predominance of
the fast solar wind found by Ulysses during its polar passages
(Phillips et al. 1995), these results led Woo and Habbal (1997)
to conclude that the fast solar wind propagates along raylike structures
originating not only from polar coronal holes but also from the quiet Sun.

The advent of the Galileo solar conjunction in January 1997 and
the capabilities of the UVCS (Kohl et al. 1997) offered the first
opportunity to test the recent interpretations of the radio measurements.
We present in this Letter the first such coordinated
measurements. We show how these simultaneous
observations provide the first direct confirmation that the slow
solar wind is limited to the streamer stalks, while the fast
wind fills the rest of the heliosphere.

\section{Characteristics of UVCS Observations}

UVCS on SOHO has proven to be a powerful tool
for probing the physical conditions in the inner corona
with measurements extending to at least 3.5 $\rm R_s$ in
coronal holes, and to 10 $\rm R_s$ in denser coronal plasmas
(Kohl et al. 1995, 1997; Noci et al. 1997).
One of the unique advantages of this instrument
is the measurement of coronal spectral lines,
in particular doublets, such as the O VI 1032 \AA\ and 1037.6 \AA\ lines,
formed primarily by the resonance scattering of chromospheric or transition
region radiation by ions flowing in the corona.
Collisional excitation also contributes to the formation of these lines.
As described in detail by Kohl $\&$ Withbroe (1982),
Noci, Kohl, $\&$ Withbroe (1987), and Withbroe et al. (1982),
the Doppler dimming effect (Hyder and Lytes 1970) is used
as a diagnostic to place limits on solar wind velocities.
As ions flow outwards in the corona, the fraction of the spectral
line formed by resonance scattering becomes Doppler-shifted
out of resonance with the disk emission.
Subsequently, the relative ratio between the intensity of the lines
forming a doublet changes drastically. 
Inferences of flow speeds from the ratio
are model dependent since they are influenced by
the electron density and
the component of the velocity distribution in the
direction of the incoming radiation to be scattered in the corona
(Noci, Kohl, $\&$ Withbroe 1987).

In the case of the O VI lines, the ratio 1032/1037 equals 4
when resonant scattering is dominant,
and reduces to 2 when only the collisional components are left.
However, this ratio can further decrease as the flow speed 
increases because of the presence of a chromospheric C II line
at 1037 \AA\ . This line is redshifted by the flow and can 
excite the O VI 1037.6 \AA\ line,
thus leading to a ``pumping'' effect (Noci, Kohl, $\&$
Withbroe 1987). 
The Doppler dimming and pumping effects have been demonstrated
successfully with UVCS in a number of coronal hole observations
(Kohl et al. 1997).

The ratio 1032/1037 of 2
occurs for a flow speed of 94 km/s, corresponding to the midpoint
separation (i.e. 0.3 \AA\ ) of the 1037.6 \AA\ and 1037 \AA\ lines,
before the onset of pumping.
A minimum in this ratio occurs for an ion flow speed of 180 km/s
when the C II line in turn 
is centered on the O VI 1037.6 \AA\ line and the pumping is at its maximum.
The ratio 1032/1037 = 2 depends only on the
velocity distribution along the flow direction.
However, the fact that the ratio 1032/1037 in our data decreases below
2 and then increases, implies that the width of 
the velocity distribution along the flow direction
is less than 0.6 \AA\ otherwise,
if the distribution along the flow direction were broader, the effect of
the C II pumping would not be detected and the ratio would not decrease below 2.
Hence, the ratio 2 used in this paper is a model independent diagnostic
for an oxygen ion flow speed of 94 km/s. While the minimum in the ratio
corresponding to 180 km/s is also model independent, it does not
provide a wide enough mapping of the flow speed in the inner corona
for the regions of interest considered in this paper.

Although the oxygen ion flow speeds could be different from
the proton/electron velocity, they are still valid proxies for
fast versus slow solar wind. Indeed multifluid solar
wind model computations show that the flow speed of minor
ions, protons and electrons are very close in
values in the inner corona (Li et al. 1997).

\section{UVCS Observations}

Taking advantage of the diagnostic tool
offered by the ratio of the oxygen lines,
as well as the contrasting signatures of the fast and slow
solar wind in Doppler scintillation
measurements, coordinated radio occultation and UVCS observations 
were carried out for the first time during the Galileo spacecraft
solar conjunction between 17 and 20 January 1997.
Figure 1 shows an image of the corona taken with the white light
coronagraph (LASCO) (Brueckner et al. 1995)
on SOHO on 17 January during this observing period. 
The slit positions of the UVCS detector were chosen to coincide
with the passage of the radio signal from Galileo through the corona
starting on 17 January.
The south polar coronal hole measurements on 19 January
were made by UVCS alone when Galileo was occulted by the
Sun. The measurements off the west limb on 20 January
followed Galileo on the egress.
To further map the solar wind velocity around a streamer,
a second set of UVCS measurements were taken on 23, 25, and 27 April 1997.
These were centered on the axis of a well-isolated streamer on the
west limb at position angle PA = 267\arcdeg\ (measured counterclockwise
from heliographic north),
as well as at 20\arcdeg\ and 40\arcdeg\ north of that position (see Figure 2).

By measuring the intensity of the two oxygen lines and their ratio along the
slits for different heliocentric distances, contours of the intensity
ratio equal 2 (or equivalently an ion flow speed of 94 km/s)
for the two observation sets were obtained. These are shown
as white lines in Figures 1 and 2. UVCS observations away from
the axes of the streamers in both data sets offered
the first direct evidence for the sharp transition
in flow speed near the boundary of the streamer to the ambient corona
at any given heliocentric distance.
A typical example illustrating this transition
is shown in Figure 3 for the UVCS observations of 27 April
at 3.5 $\rm R_s$ at 20\arcdeg\ north of the streamer stalk shown in Figure 2.
Figure 3 clearly illustrates how the relative change in peak
intensity and shape of the two oxygen line profiles 
varies significantly as a function of latitude, or more relevantly,
as a function of position angle away from the streamer stalk.

Plots of the ratio of the two oxygen lines
versus heliocentric distance
for different position angles SA =  0\arcdeg\ , $\pm$ 10\arcdeg\ and
$\pm$ 20\arcdeg\
measured north (+) or south (-) with respect to the streamer stalks
are shown in Figure 4. The top three panels correspond to the
January east limb observations, and the lower four panels to the
April observations.
For small heliocentric distances the data were averaged over 2-4'.
At distances larger than 3.5 $\rm R_s$ the data were averaged over 11'
(corresponding to an uncertainty of 2\arcdeg\ )
when the streamer stalk was not in the field of view.
It is clear from these plots that the wind reaches a speed of 94 km/s
(or ratio = 2) around 4.5 $\rm R_s$ along the streamer stalk (SA =
0\arcdeg\ ). In contrast, the wind is faster closer to the Sun as it
moves away from the streamer stalk, for example, at 3.5 - 4
$\rm R_s$ for SA = $\pm$ 10\arcdeg\ ,
or 2.5 - 3 $\rm R_s$ for SA = $\pm$ 20\arcdeg\ .
The first minima seen at $\pm$ 20\arcdeg\ are very close
in heliocentric distances to those found in coronal holes,
and correspond to 180 km/s. An uncertainty of 0.5 in the ratio
corresponds to an uncertainty of 25 km/s in speed.

Most striking in the ratio = 2 contours is the sharp latitudinal
gradient in wind speed that occurs close to the stalk of the streamers.
However, not only is the change in ratio indicative of changes in solar wind
character, but so is the width of the spectral lines. As shown in Figure 3,
the lower ratio and broader profile are typical of a fast and hot
wind (as far as the oxygen ions are concerned). These are comparable to UVCS
measurements in polar coronal holes, and are
a strong indication of the anisotropy
of the velocity distribution in the fast wind as shown by Kohl et al. (1997).
Since the line ratios are derived from measurements
along the line of sight, the flow speed is high over a large
fraction of the line of sight. Along the
streamer stalk, on the other hand, the ratio is higher, and the lines
narrower, an indication of cooler and slower flowing ions,
as reported in earlier UVCS observations of streamers
(Noci et al. 1997).

\section{Radio Occultation Measurements}

The corresponding Doppler scintillation measurements made by Galileo
from 15 January to 4 February are shown in Figure 5.
These measurements are normalized to
heliocentric distance assuming a 1/$r^2$ dependence in density
fluctuation (Woo $\&$ Gazis 1993). They are characterized by
relatively low levels of Doppler scintillation except on 22 January.  The
white light image of 22 January in Figure 6 shows that the enhanced Doppler
scintillation on 22 January is caused by a streamer stalk intercepting
the Galileo radio path, thus confirming that pronounced enhancements
observed in Doppler scintillation in the absence of coronal mass ejections
(Woo 1997) are a proxy for coronal streamer stalks (Woo et al. 1995).
Results from these coordinated measurements also reinforce those obtained
from comparing intensity scintillation
and solar wind speed measurements (Woo $\&$ Martin 1997), i.e., low levels
of Doppler scintillation observed away from streamer stalks are a
proxy for the fast wind.  The use of Doppler scintillation as a proxy for
wind speed is particularly important because its high sensitivity to
changes in electron density over small distances makes it a measurement of
high spatial resolution (Woo 1996).

\section{Discussion and Conclusion}

Radio occultation measurements provided the
first hints of the source regions of the fast and slow solar
wind and the impetus for the present coordinated observations.
The results reported here illustrate the
power of the UVCS instrument to provide more definitive answers,
since it can map the whole plane of the sky.

Although a number of UVCS observations of streamers have been made
since the launch of SOHO (see, e.g., Noci et al. 1997),
the coordinated radio scintillation and ultraviolet observations
of the inner corona presented here are the first of their kind.
The most straightforward result to emerge from these coordinated
observations is that low levels of radio scintillation are indeed
associated with fast solar wind, while the pronounced peaks result from
the passage of the radio signal through the streamer stalks.
More importantly however, this work provides the first
confirmation of the perspective recently developed
from radio occultation measurements (Woo 1995; Woo $\&$ Habbal
1997; Woo $\&$ Martin 1997), namely
that the streamer stalks are the locus of the slowest solar
wind, while the fast solar wind dominates the corona.

These coordinated observations also confirm the existence
of sharp gradients in solar wind speed
found earlier by radio occultation measurements of the corona (Woo
1995), and show that they occur along the streamer stalks.
Undoubtedly, remnants of this velocity shear must survive in interplanetary space 
since they are frequently observed in in situ observations beyond
0.3 AU (e.g. Schwenn et al. 1978; Rhodes and Smith 1981).

The UVCS observations presented here also support
the view recently proposed by Woo and Habbal (1997) that the fast solar
wind does not necessarily originate only from polar coronal holes;
its ubiquitous nature, so vividly evident in the Ulysses
measurements, can also derive from its origin in the quiet Sun
regions. The fraction of the fast solar wind
originating from these latter regions, however, cannot be inferred from the
present observations. 

That there exist two types of solar wind, namely the fast and slow,
with different physical characteristics can be readily understood
if we consider their corresponding magnetic sources.
It seems plausible that radially extending raylike structures
originate within the boundaries of supergranular cells which
indiscriminately cover the solar surface in coronal holes and the
quiet Sun (Title 1997).
These cells are preserved in coronal holes because of the absence
of large scale closed magnetic field lines. In the quiet Sun,
the supergranular cells at coronal heights are essentially
preserved except for occasional disruptions by
large scale magnetic field lines
interconnecting widely separated magnetic regions
and forming its diffuse characteristic emission.
Indeed, close inspection of eclipse observations of the Sun
clearly show the simultaneous presence
of open and closed magnetic structures along the line of sight at low
latitudes (e.g. Koutchmy 1977; November $\&$ Koutchmy 1996).
On the other hand,
the streamer stalks which carry the slow solar wind
belong to the large scale coronal structures which have dominated our
impression of the corona for so long and which derive from deep-rooted
multipolar fields.

The new clues provided by the results of this study should
lead to new perspectives in the search for the elusive
coronal heating mechanisms of the solar wind. In particular, if indeed
a fraction of
the fast solar wind also originates from quiet regions, then the energy
flux per unit area requirements to the solar wind from its sources at the Sun
can be significantly reduced.

\acknowledgements

We thank the UVCS team for the tremendous efforts expended in producing
such a unique instrument, and the Galileo Project, the JPL Radio Science
Systems Group, and the NASA Deep Space Network (DSN) for conducting the
radio measurements. Special thanks are due
to Cindy Copeland for her preparation of Figures 1 and 2.
This research was supported by the National Aeronautics and Space
Administration (NASA) under grants NAGW-4381 and NAGW-4993 (S. R. Habbal)
and NAG5-3192 to the Smithsonian Astrophysical Observatory,
by Agenzia Spaziale Italiana, and by Swiss funding agencies,
and under a contract with NASA to the Jet Propulsion Laboratory,
California Institute of Technology.
LASCO was constructed and is operated by an international 
consortium consisting of the Naval Research Laboratory (Washington,
DC), the Department of Space Research at the University of Birmingham
(Birmingham, United 
Kingdom), the Max-Planck Institute for Aeronomy (Lindau, Germany) and
the Laboratory for Space Astronomy (Marseilles, France).  SOHO is a
mission of international cooperation between ESA and NASA.

\newpage

\figcaption[FIG1]{
White light image of the corona taken with the LASCO
C2 coronagraph on SOHO on 17 January 1997. The bright object in the east below
the equator is Jupiter and indicates the approximate location of
the Galileo spacecraft. The field of view spans 2 to 6 $\rm R_s$. 
The spatial length of the field of view defined by the
slit of the UVCS detector is approximately 2 $\rm R_s$.
Shown as black vertical lines are the slit positions,
located at 1.9, 2.5, 3, 4, 4.7 and 5.5 $\rm R_s$ on
the east limb, at 1.9 $\rm R_s$ in the south polar coronal hole, and
at 4, 4.7 and 5.5 $\rm R_s$ at the west limb.
They are perpendicular to the radial direction
at position angles PA = 97\arcdeg\ ,
180\arcdeg\ and 247\arcdeg\ respectively,
measured counterclockwise from heliographic north.
The white contours mark the ratio of the oxygen 1032/1037 
line intensities equal to 2, or, equivalently 94 km/s. \label{1}}

\figcaption[FI2]{
Same as Figure 1 for observations on 23, 25 and 27 April, 1997.
The background white light corona from LASCO was taken on 27 April.
These observations were made with slit positions perpendicular
to the axis of the streamer at PA = 267\arcdeg\ at 2.3, 3., 3.5, 4., 4.5,
and 5 $\rm R_s$. 
Additional observations were made at 20\arcdeg\ north 
of these positions and one at 40\arcdeg\ north of the position
at 3 $\rm R_s$.
Here too the white contours mark the ratio of the oxygen 1032/1037 
line intensities equal to 2. \label{2}}

\figcaption[FIG3]{
Top panel: False-color image of the intensity 
of the O VI 1032 \AA\ , 1037.6  \AA\ and Ly $\beta$ lines
along the detector slit (or spatial direction/vertical axis) positioned
at 3.5 $\rm R_s$, 20\arcdeg\ north of the streamer axis of
Figure 2. The horizontal axis is the spectral direction.
The resolution of a bin element is 0.28 \AA\ in the spectral
direction (1 bin = 2 detector pixels)
and 28'' in the spatial direction. Tick marks are
spaced approximately every 1 \AA\ and 1' in the spatial and spectral
directions respectively. Note that
because of the pointing on 27 April,
the roll angle of UVCS is such that north faces downward in this figure.
Bottom panels: Profiles of the 1032 \AA\ (left) and 1037.6 \AA\ (right)
lines averaged over (a) 4.5', (b) 5.25'
and (c) 11.5', as indicated
by the corresponding labeled spaces between the arrows in the top
panel. The scale on the horizontal axis is in detector pixel (or 0.14
\AA\ ).
The ratio of the line intensities is (a) 2.1 $\pm$ 0.1,
(b) 1.7 $\pm$ 0.25 and (c) 1.4 $\pm$ 0.5 respectively,
indicating an increase in flow speed from (a) to (c). \label{3}}

\figcaption[FIG4]{
Plots of the ratio of the two oxygen lines
versus heliocentric distance R/$\rm R_s$
for different position angles SA =  0\arcdeg\ , $\pm$ 10\arcdeg\ and
$\pm$ 20\arcdeg\ measured north (+) or south (-)
with respect to the streamer stalk.
Panels (a)-(c) correspond to the
January east limb observations, and (d)-(g) refer to the
April observations. \label{4}}

\figcaption[FIG5]{
Radio Doppler scintillation measurements by Galileo
during its solar conjunction from 15 January to 4 February 1997, or
day of year (doy) 15 to 35. \label{5}}

\figcaption[FIG6]{
White light image on 22 January 1997 from the LASCO C3 coronagraph,
with a field of view extending from 3.7 to 30 $\rm R_s$.
Jupiter (the bright object) indicates the point of closest approach
of the line of sight radio path from Galileo. \label{6}}


\begin{references}

\reference{bell1975} Bell, B. $\&$ Noci, G. 1976
\jgr, 81, 4508

\reference{brueck1995} Brueckner, G. E. et al. 1995, \solphys, 162, 357

\reference{gosl1997} Gosling, J.T. 1997
in {it Robotic Exploration Close to the Sun: Scientific Basis},
S. R. Habbal (Ed.), AIP CP-385, p. 17

\reference{gosl1995} Gosling, J.T. et al. 1995, \grl, 22, 3329

\reference{harv1996} Harvey, K. L. 1996, 
in {\it Solar Wind Eight}, D. Winterhalter, J. T. Gosling,
S. R. Habbal, W. S. Kurth and M. Neugebauer (Eds.), AIP CP-382, p. 9

\reference{hund1977} Hundhausen, A.J. 1977, 
in {\it Coronal Holes and High Speed Wind Streams}
(ed. Zirker, J.B.) 225 (Colo. Assoc. Univ. Press, Boulder)

\reference{hyd1970} Hyder, C. L., $\&$ Lites, B. W. 1970,
\solphys, 14, 147.

\reference{kohl1982} Kohl, J. L., $\&$ Withbroe, G. L. 1982,
\apj, 256, 263

\reference{kohl1995} Kohl, J. L. et al. 1995,  \solphys, 162, 313

\reference{kohl1997} Kohl, J. L. et al. 1997, 
\solphys (in press)

\reference{kout1977} Koutchmy, S. 1977, \solphys, 51, 399.

\reference{krieg1973} Krieger, A. S., Timothy, A. F., $\&$  Roelof, E. C. 1973,
\solphys, 23, 123

\reference{li1997} Li, X., Esser, R., Habbal, S. R., $\&$  Hu, Y.-Q. 1997,
\jgr, 102, 17419

\reference{munro1977} Munro, R. H. $\&$ Jackson, B. V. 1977,
\apj, 213, 874

\reference{noci1987} Noci, G., Kohl, J. L., $\&$  Withbroe,
G. L. 1987,
\apj, 315, 706

\reference{noci1997} Noci, G., et al. 1997, 
Advances in Space Res. (in press)

\reference{nov1996} November, L., $\&$  Koutchmy, K. 1996, \apj, 466, 512

\reference{phil1995} Phillips, J. et al. 1995,
\grl, 22, 3301

\reference{rhod1981} Rhodes, E., Jr., $\&$  Smith E. 1981,
\jgr, 86, 8877

\reference{schw1990} 
Schwenn, R. 1990, in {\it Physics of the Inner Heliosphere:
1. Large-Scale Phenomena}, R. Schwenn and E. Marsch (eds.),
p. 99, Springer-Verlag

\reference{schw1978} Schwenn, R.,  et al. 1978,
\jgr, 83, 1011

\reference{titl1997} Title, A. M. 1997,
EOS, 78, S243

\reference{with1982} 
Withbroe, G. L., Kohl, J. L., Weiser, H., $\&$  Munro, R. H. 1982,
Space Sci. Rev., 33, 17

\reference{woo1995} Woo, R. 1995,
\grl, 22, 1393

\reference{woo1996} Woo, R., 1996,
in {\it Solar Wind Eight}, D. Winterhalter, J. Gosling, S. Habbal,
W. Kurth, and M. Neugebauer (eds.), AIP CP-382, New York, 38

\reference{woo1997a} 
Woo, R. 1997, in {\it Coronal Mass Ejections: Causes and Consequences},
N. Crooker, J. Joselyn and J. Feynman (Eds.),
AGU Monograph 99, AGU, Washington D.C., in press.

\reference{woo1997b} Woo, R. $\&$  Habbal, S.R. 1997,
\grl, 24, 1159

\reference{woo1993} Woo, R. $\&$  Gazis, P. 1993,
\nat, 366, 543

\reference{woo1997c} Woo, R. $\&$  Martin, J. 1997,
\grl, submitted

\reference{woo1995b}
Woo, R., Armstrong, J.W., Bird, M.K., $\&$ P\"atzold, M. 1995,
\apj, 449, L91

\end{references}
\end{document}